\documentclass[twocolumn,showpacs,showkeys,preprintnumbers,amsmath,amssymb,prl]{revtex4}
\usepackage{graphicx}% Include figure files
\usepackage{dcolumn}% Align table columns on decimal point
\usepackage{bm}% bold math
\usepackage{ulem}% cross out
\begin{document}

\title{Optomechanical Stochastic Resonance in a Macroscopic Torsion Oscillator}

\author{F. Mueller}
\author{S. Heugel}
\author{L. J. Wang} 
\email{lwan@optik.uni-erlangen.de}
\affiliation{Max Planck Research Group, Institute of Optics, Information and 
Photonics, University Erlangen-Nuremberg, D-91058 Erlangen, Germany}
\homepage{http://www.optik.uni-erlangen.de}

\begin{abstract}
Linear mechanical oscillators have been applied to measure very small forces, mostly with the help of noise suppression. In contrast, adding noise to non-linear oscillators can improve the measurement conditions.
Here, this effect of stochastic resonance is demonstrated in a macroscopic torsion oscillator, for an optomechanical non-linear potential. The signal output is enhanced for a sub-threshold electronic signal. This non-linear oscillator serves as a model system for the enhancement of signal-to-noise ratio in high precision optomechanical experiments.
\end{abstract}

\maketitle
% ************************ NEW BEGIN ************************

% ************************ INTRO ****************************
The measurement of very weak forces~\cite{brag2_1972}, especially those of a gravitational nature, has great importance as it quite often tests our basic understanding of the fundamental physical laws. In such measurement systems, noise is almost universally an unwanted effect as it imposes a limitation of the measurement sensitivity and precision. Various noise types have to be considered especially in high precision, weak force measurements. Prominent examples include gravitational experiments~\cite{brag2_1972,baessler1999}, the measurement of weak radiation pressure effects~\cite{rohrbach2005,mueller2_2008}, and combined optomechanical systems~\cite{meystre1985,kleckner2006}.

A typical example for a measurement system with extremely high precision, in which the effect of radiation pressure is non-negligible, is a gravitational wave detector~\cite{harms2003}. Here, the combined optomechanical system shows different characteristics and new physical phenomena such as the optical spring effect and multistability~\cite{mueller3_2008} become important.
While the fundamental limits for noise reduction in linear systems are well characterized, both theoretically and experimentally~\cite{gillies1992,mueller2007}, a totally different picture emerges for a combined, non-linear, optomechanical multistable system.
In such a system, stochastic resonance (SR)~\cite{benzi1983,badzey2005} may appear where by  adding stochastic noise one can enhance the signal-to-noise ratio (SNR) in precision measurements. Furthermore, if one can synchronize the system's natural inter-well transition rate, the so-called Kramers rate~\cite{kramers1940}, to the frequency of a very weak, sub-threshold harmonic signal, this may lead to an increased system response~\cite{gamma_1998}. The effect of SR has been observed in a few systems, such as bistable ring lasers, semiconductor devices, nano-mechanical devices, and neuronal physiological spiking processes in cells~\cite{badzey2005,mcnamara1988,douglass1993}.

Here, we report the observation of stochastic resonance in a macroscopic, optomechanically coupled oscillator. We implemented a thermal-noise limited torsion oscillator that is sensitive to femto-Newton level weak forces~\cite{mueller2_2008}. Furthermore, the oscillator is coupled to an optical cavity, and its dynamics now follow an asymmetric multistable optomechanical potential~\cite{mueller3_2008}. We apply a weak electronic signal and observe an SR-like output enhancement. This signal amplification process agrees well to a theoretical model but also shows major deviations from the predictions given for a symmetric bistable potential.
% ************************ INTRO END ************************

% ************************ THEORY ***************************
A linear system yields equal SNR for the input and the output signal~\cite{mcnamara1989}.
In contrast, an SR system exhibits increased output SNR for an increased amount of input noise.
This counterintuitive concept can be understood by considering an overdamped oscillator in a double-well potential.
For a torsion balance oscillator with angular position variable $\varphi$, the equation of motion reads
\begin{equation}
\label{equ1}
2I\gamma^{*}\dot{\varphi}-a\varphi^3+b\varphi=T_{S}\cos{\omega_{S} t}+I\alpha(t),
\end{equation}
when overdamping ($\ddot{\varphi}=0$) is assumed.
$T_{S}$ is the amplitude of an external signal torque, modulated at a frequency $\omega_{S}$, and $\alpha(t)$ represents a time-dependent angular acceleration due to added thermal noise~\cite{mueller2007}.
The total damping rate $\gamma^{*}$ is the system's friction parameter, and $I$ the moment of inertia.
Parameters $a$ and $b$ represent the system's scaled potential parameters.
Such an oscillator's static bistable potential $U(\varphi)$ is
\begin{equation}
\label{equ2}
U(\varphi)=\frac{a}{4}\varphi^4-\frac{b}{2}\varphi^2,
\end{equation}
with two angular position minima at $\pm\varphi_m=\pm\sqrt{b/a}$, separated by a potential energy barrier of height $\Delta U=b^2/4a$.
Unless the total noise power is much smaller than the given potential barrier, the oscillator seeks stable angular positions around $\pm\varphi_m$, while the additional noise fluctuation torque $I\alpha(t)$ causes occasional transitions of the oscillator between the two position minima.
With increasing noise, the probability of noise-driven transitions increases, quantified by the Kramers rate~\cite{gamma_1998,kramers1940}
\begin{equation}
\label{equ3}
r_K=\frac{\omega_m\omega_b}{2\pi\gamma^{*}}e^{-\frac{\Delta U}{U_D}},
\end{equation}
which gives the average rate of inter-well transitions in a double-well potential with a smooth curvature at the energy barrier.
Here, $\omega_m^2=U''(\varphi_m)/I$ and $\omega_b^2=\left|U''(\varphi_b)/I\right|$ are the squared mechanical frequencies at the potential minima $\pm\varphi_m$ and at the potential barrier $\varphi_b$, respectively. 
$U''(\varphi)$ is the system's total local torsion constant.
The value $U_D=k_B T$ is the noise energy in the system, calculated as the equivalent noise temperature at the system's \lq\lq temperature\rq\rq~$T$, with the Boltzmann constant $k_B$~\cite{mueller2007}.

The noise-dependent SNR for the system response in the presence of the external modulation~\cite{gamma_1998,mcnamara1989,wiesenfeld1995} is
\begin{equation}
\label{equ4}
SNR=\pi\left(\frac{U_S}{U_D}\right)^2 r_K\propto\left(\frac{U_S}{U_D}\right)^2 e^{-\frac{\Delta U}{U_D}}.
\end{equation}
This expression implies a signal energy $U_S=T_S\varphi_m$ smaller than the bistable potential's energy barrier $\Delta U$, and in addition requires the adiabatic approximation $\omega_S<<\omega_m$, which means that the system returns to its equilibrium position much faster than a change in the external modulation signal occurs.
$SNR$ reaches a maximum for a noise level $\overline{U}_D=\Delta U/2$.
At this level, the inter-well transition rate is synchronized with the external modulation, such that the Kramers rate is twice the external signal frequency, $r_K=\omega_S/\pi$.

A common way to quantify $SNR$ is to calculate the power spectral density of the system's signal output, compared to the spectral background, both evaluated at the input signal frequency~\cite{badzey2005,douglass1993}. 
However, for the multistable system used here, a residence time analysis method~\cite{gamma_1998} is more appropriate because it pre-filters irrelevant intra-well dynamics from the data, using a threshold filter. This gives a discrete position signal. 
For analysis, the residence times in either potential level are binned, and SR is quantified~\cite{gamma_1998} by calculating the total number of inter-well transitions around the signal's half period $T/2=\pi/\omega_S$, for changing noise energy levels.
The evaluation of such transitions gives a signal strength, proportional to $SNR$.
Higher order odd harmonic contributions can be added as signal contributions, e.g.~for square wave excitation~\cite{casado-pascual2004,badzey2005}.
The discretization method can be expanded for the multistable optomechanical potential by considering $n$ discrete levels, where the filter threshold levels are determined from the positions with least residence probability.

% ************************ THEORY END ***********************

% ************************ SETUP ****************************
\begin{figure}                                          
\includegraphics[width=8.2cm]{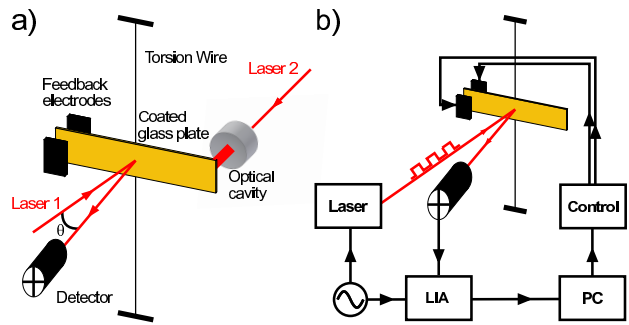}
\caption{Experimental scheme. The torsion balance 
oscillator (a) is a precision force measurement device, 
sensitive down to the fN range. Linear and non-linear control techniques 
exist for the applied electrostatic feedback (b). The stable optical coupling into 
a hemi-spherical cavity is feasible due to the torsion oscillator's horizontal alignment. 
More details in the text. LIA: Lock-in amplifier.}
\label{fig1}
\end{figure}
The experimental system is shown schematically in Fig.~\ref{fig1}. 
It consists of a torsional oscillator~\cite{gillies1992} made of a 
gold-coated glass plate, $50\,mm\times10\,mm\times0.15\,mm$ in size, 
doubly suspended on a $15\,cm$ long, $25\,\mu m$ diameter tungsten wire. 
The oscillator body has a mass of $\sim0.2\,g$ and a moment of 
inertia $I=4.6\,\times\,10^{-8}\,kg\,m^{2}$. 
The measured torsion constant is $\tau=2.2\times10^{-7}\,Nm\,rad^{-1}$. 
The torsion pendulum has a natural frequency of $f_{0}=0.36\,Hz$ 
with a quality factor $Q\sim2,600$.

A laser beam is reflected from the center of the oscillator and 
detected by a high-sensitivity quadrant diode detector followed 
by a lock-in detector~\cite{lorrain1991}. 
The oscillator's angular position voltage signal is digitized at 
a sampling rate of $5\,kHz$.  
This measurement scheme has an angular position sensitivity of $2\,nrad\,Hz^{-1/2}$, which equals a
linear displacement of the oscillator arm of $0.4$ \rm{\AA} with respect to the feedback electrodes.
Subsequently, the signal is used as the input of a computerized, digital control loop, which allows to generate proportional and differential control schemes. 
The differential scheme enables active derivative damping of the oscillator.
The control signal is converted to an analog output signal applied to two electronic feedback electrodes, which enables efficient control of the balance's dynamics~\cite{mueller_diss2008}.
As an example, the oscillator's total friction $\gamma^{*}$ can be artificially adjusted, and by fine tuning the relevant parameter, the system's noise energy is controlled. Although this implies a dependence between friction and noise, the method is found to give precisely adjustable noise levels while the total friction only changes slightly.

For the generation of optomechanical coupling, the oscillator body's gold-coated glass plate serves as the moving flat mirror of a hemi-spherical optical cavity~\cite{mueller3_2008}. 
A second, spherical mirror with a curvature radius of $25\,mm$ is rigidly mounted opposite to the glass plate, at a distance of $12.5\,mm$.
When a second laser with a wavelength of $\sim 660\,nm$ is coupled in, this cavity forms Laguerre-Gaussian TEM$_{00}$ and TEM$_{20}$ modes with a free spectral range of the fundamental mode at $\sim13.5\,GHz$.
The optical cavity has a low finesse of $F=11$, giving a mean mirror reflectivity of $R=0.87$.
The oscillator's absolute measurement sensitivity is $100\,fN$ or $15\,\mu W$ of optical power for the detection of radiation pressure in total reflection~\cite{mueller2_2008}.
A high vacuum ($10^{-7}\,mbar$) environment encloses the setup, which itself is mounted on top of an active vibration isolation system.
% ************************ SETUP END ***************************

% ************************ MEAS - ELECTRONIC *******************
\begin{figure}[htbp]                                         
\includegraphics[width=6cm]{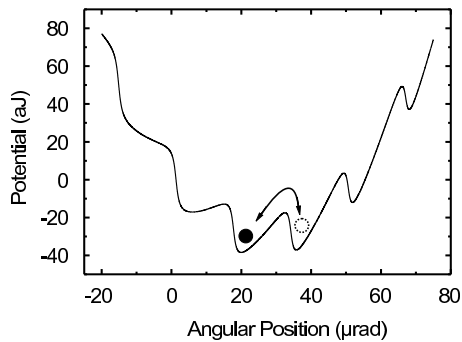}
\caption{The optomechanical potential is generated with an optical cavity input power of $32\,mW$. Stable potential minima are formed by TEM$_{00}$ cavity modes (higher orders not considered), with an angular spacing of $\sim 16\,\mu rad$. The potential plot shows the directional asymmetry, with an average potential depth of the centered wells of $\sim 20\,aJ$.
This is also the smallest applied signal energy if the oscillator is to be moved directly from one potential minimum to the next.}
\label{fig2}
\end{figure}
The optomechanical potential (shown in Fig.~\ref{fig2}) exhibits several non-linear effects, such as angular position multistability and hysteresis~\cite{meystre1985,corbitt04_2007,mueller3_2008}.
This potential is used for a measurement of optomechanical SR.
Experimentally, the mechanical torsion constant of the free system is electronically lowered to $\tau=9.6\times10^{-8}\,Nm/rad$, equal to a mechanical oscillation period of $T_0=4.3\,s$. 
This makes the additional optical potential dominant.
Then, the torsion balance is optomechanically coupled using a cavity optical input power of $P_{in}=32\,mW$.
The two centered TEM$_{00}$ mode potential minima (at $\sim 20\,\mu rad$ and $\sim 36\,\mu rad$) have an average potential depth of $\sim 20\,aJ$.
The optical spring effect~\cite{mueller3_2008} increases the local optomechanical torsion constant to $\tau_{os}=2.2 \times 10^{-6}\,Nm/rad$, equal to an oscillation period of $0.9\,s$.
This \lq\lq optical spring\rq\rq~constant slightly increases from one potential minimum to the next, due to the local potential curvature (see Fig.~\ref{fig2}).
Here, the weak modulation signal is an electronic square wave applied to the feedback electrodes, with a frequency $\omega_S=2\pi\times 100\,mHz$ and a torque amplitude of $T_S=0.79\,pNm$.
This gives a sub-threshold signal energy $U_S=T_S\varphi_m=6.2\,aJ$.

\begin{figure}[htbp]                                         
\includegraphics[width=8cm]{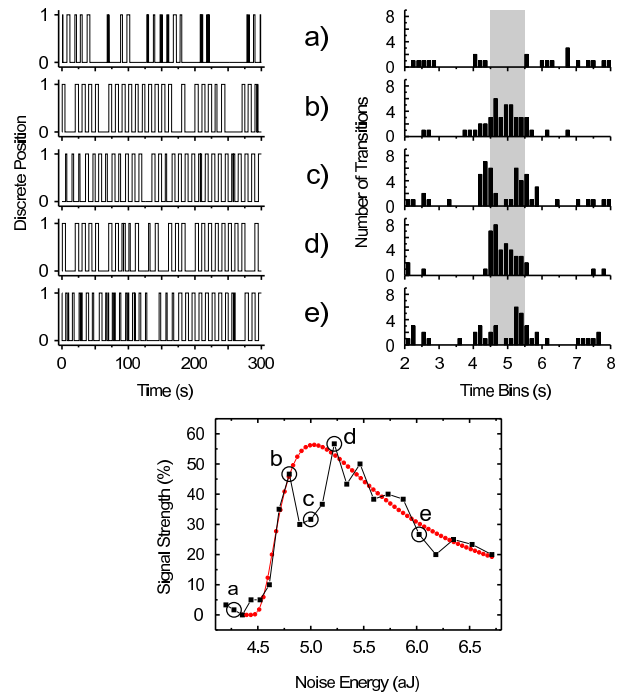}
\caption{SR in the optomechanical potential. The multistable system response is shown (upper left plot) for an applied weak electronic modulation signal with a signal energy $U_S=6.2\,aJ$, at a frequency $\omega_S=2\pi\times 100\,mHz$. The noise energy varies from  $U_D=4.4\,aJ$ (a), $U_D=4.8\,aJ$ (b), $U_D=5.0\,aJ$ (c), $U_D=5.2\,aJ$ (d), to $U_D=6.0\,aJ$ (e). For the discrete filtering (upper left), two neighboring TEM$_{00}$ optical cavity modes are considered. 
Time bin histograms (upper right plot) serve to find the degree of coherence between the excitation signal and the transition rate. SR is quantified from the total number of transitions, proportional to the signal strength. The relevant transitions are shown in the shaded region around $T/2=(5\pm0.5)\,s$.
The black curve in the lower plot shows the signal strength for the total measurement. It exhibits two maxima at $\overline{U}_{D1}=4.8\,aJ$ and $\overline{U}_{D2}=5.2\,aJ$, with a clear minimum around $U_D=5.0\,aJ$. Ignoring the local minimum, a fit (red curve) to the theoretical model of Eq.(\ref{equ4}) is also shown. This curve's maximum gives an exponential factor $e^{-\frac{\Delta U}{U_D}}=e^{-1.98}$, which agrees to the theoretical curve maximum at $\overline{U}_D=\Delta U/2$. Thus, experiment and theory are in good agreement for this coarse resolution.
However, a fine resolution leads to SR splitting around (c), implying a dependence of SR on the applied potential's local shape (more details in the text).}
\label{fig3}
\end{figure}

The upper left plot of Fig.~\ref{fig3} shows five discrete data sets of $300\,s$ each, with applied modulation signal, for a noise energy ranging from $U_D=4.4\,aJ$ (a) to $U_D=6.0\,aJ$ (e).
If the system's response is fully coherent with the driving force, a maximum of $60$ transitions should occur during a single measurement.
The filtering process considers only the two centered TEM$_{00}$ modes. 
The angular position sign is inverted with respect to the potential simulation shown in Fig.~\ref{fig2}.
As the noise energy increases, the degree of signal coherence increases, but the system output is never fully coherent to the excitation signal.
For lower noise levels (a,b), the oscillator appears to have a higher residence probability in the lower discrete position, which means that it \lq\lq prefers\rq\rq~the smoother transition.
For higher noise levels (d,e), it also follows the steeper transition more frequently.

Furthermore, the residence time bins are calculated with a width of $0.15\,s$, and a signal bin centered at the excitation signal's half period $T/2=5\,s$.
The signal strength is evaluated $\pm 0.5\,s$ around $T/2$.
This signal range considers the numerical error of the discrete filtering process.
The upper right plot of Fig.~\ref{fig3} shows the residence time bins for the given measurements, where the shaded region represents the signal time bins.
The lower plot of Fig.~\ref{fig3} shows the scaled signal strength as a function of noise energy.
Here, the black curve represents all measured inter-well transitions, independent of their direction.
A clear minimum is found in this curve at a noise energy of $U_D=5.0\,aJ$ (c), between two maxima at $\overline{U}_{D1}=4.8\,aJ$ (b) and $\overline{U}_{D2}=5.2\,aJ$ (d).
Ignoring the datapoints between (b) and (d) in a first analysis step, a coarse fit (red curve) according to the model given in Eq.(\ref{equ4}) nicely agrees to the measurement, also giving a maximum at $\overline{U}_D=\Delta U/2$ (see caption of Fig.~\ref{fig3}).

The fit's result corresponds well to an assumed symmetric optomechanical potential.
However, the refined analysis (datapoints around (c)) reveals a deviation from this assumption, explained in the following.

The energy thresholds for two neighboring potential wells show a slight difference, resulting in a small transition rate splitting~\cite{gamma_1998}. 
In addition, the distinct optomechanical potential asymmetry influences the result, even for equal depths in neighboring potential wells.
The SR condition for the smoother transition appears to be fulfilled at a lower noise energy than for the steeper transition.   
Between the resonant noise levels (c), neither of the transitions is preferred.
Looking at the histogram plot (upper right of Fig.~\ref{fig3}), a clear transition rate splitting is observed for the noise levels around this central SR minimum.

Considering a direction-dependent signal strength helps to clarify these results.
Comparing the optomechanical potential's shape (see Fig.~\ref{fig2}) with the assumptions made in Eqs.(\ref{equ2}) and (\ref{equ3}) for a symmetric bistable potential, one expects slightly different values for $\omega_m$ at each minimum, and thus non-equal Kramers rates in either potential well for a constant noise energy $U_D$.
Therefore, the two considered directional transition rates are synchronized with the external excitation signal for slightly different noise energies, which leads to the observed SR splitting.
Thus, the system response to noise excitation depends on its current discrete position state.
The deviations from the symmetric theoretical model can be explained by the optomechanical potential's intrinsic intra-well and inter-well asymmetry.
% ************************ MEAS - Optomech END ****************
% ************************ CONCLUSION **************************

In summary, the torsion balance oscillator coupled to the optomechanical potential clearly shows SR phenomena for an applied weak modulated electronic force.  
The degree of coherence with a sub-threshold modulation signal is adjustable by tuning the noise energy in the system.
The free torsion balance is a linear instrument.
However, when coupled with the intracavity light pressure, the system becomes highly non-linear. Furthermore, since this new optomechanical system displays an asymmetry between adjacent potential wells, a symmetric potential theory cannot fully explain the observed optomechanical SR splitting.
This deviation is clearly observed experimentally in the form of a splitting of the SR signal enhancement ratio. We further provide a semi-qualitative explanation for our experimental results, based on the fact that the local potential properties influence the inter-well transition rates.
Further theoretical treatment is being developed to explain such effects.
Since the optomechanically coupled torsion oscillator serves as a testbed for mechanical systems coupled to a resonant light field, the application of SR methods may find further applications in high-precision optomechanical force measurements.

We thank S. Malzer and B. Menegozzi for technical help, and Z. H. Lu for helpful discussions.
% ************************ CONCLUSION END **********************

%\bibliography{StochRes}

\end{document}